\documentclass[12pt,english,floatfix,superscriptaddress,aps,prd,preprint]{revtex4}
\usepackage[utf8]{inputenc}
\usepackage{amsmath}
\usepackage{amssymb}
\usepackage{amsbsy}
\usepackage{amsfonts}
\usepackage{amsopn}
\usepackage{amstext}
\usepackage{graphicx}
\usepackage{amssymb}
\usepackage{amsfonts}
\usepackage{amsmath}
\usepackage{amsmath,amsthm,amsfonts,amssymb}
\usepackage[mathcal]{eucal}
\usepackage{mathrsfs}
\usepackage{graphicx}
\usepackage[english]{babel}
\usepackage{color}
\usepackage{esint}
\usepackage[dvips]{epsfig}
\usepackage[dvips]{graphicx}
\usepackage{float}
\usepackage{units}
\usepackage{textcomp}

\usepackage{hyperref}
\usepackage{slashed}

\newcommand{\ie}{\begin{equation}}
\newcommand{\fe}{\end{equation}}
\newcommand{\se}{\begin{eqnarray}}
\newcommand{\ff}{\end{eqnarray}}

\begin{document}

\title{Noninertial and spin effects on the 2D Dirac oscillator in the magnetic cosmic string background}


\author{R. R. S. Oliveira}
\email{rubensfq1900@gmail.com}
\affiliation{Universidade Federal do Cear\'a (UFC), Departamento de F\'isica,\\ Campus do Pici, Fortaleza - CE, C.P. 6030, 60455-760 - Brazil.}

\date{\today}

\begin{abstract}

In this work, we analyze the influence of noninertial and spin effects on the dynamics of the 2D Dirac oscillator in the magnetic cosmic string background. To model this background, we consider a uniform magnetic field, the Aharonov-Bohm effect, and a parameter $\eta$ generated by a cosmic string. Posteriorly, we determine the bound-state solutions of the system: the Dirac spinor and the relativistic energy spectrum. We verified that this spinor is written in terms of the generalized Laguerre polynomials and this spectrum depends on the effective quantum number $N_r$, angular velocity $\Omega$ and parameter $s$ associated to the noninertial and spin effects, magnetic flux $\Phi$, cyclotron frequency $\omega_c$, zero-point energy $E_0$, and on the deficit angle $\eta$. In particular, we note that besides this spectrum to be a periodic function and asymmetric, its values infinitely increase when $\eta\to 0$ or $N_r=\omega_c=\Omega\to\infty$. We also note that the energies of the antiparticle with spin down are larger than of the particle with spin up or down. In the nonrelativistic limit, we get the Schr\"{o}dinger-Pauli oscillator with two types of couplings: the spin-orbit coupling and the spin-rotation coupling, and two Hamiltonians: one quantum harmonic oscillator-type and other Zeeman-type. Finally, we compare our results with other works, where we verified that our problem generalizes some particular cases of the literature when $\Omega$, $\omega_c$, $\Phi$, $s$ or $\eta$ are excluded from the system.

\end{abstract}

\maketitle


\section{Introduction}

In 1989, M. Moshinsky and A. Szczepaniak formulated the first relativistic version of the quantum harmonic oscillator (QHO) for spin-1/2 particles, in which it became known as Dirac oscillator (DO) \cite{Moshinsky}. To configure the DO, is necessary inserted into free Dirac equation (DE) a nonminimal coupling given by: ${\bf p}\to{\bf p}-im_0\omega\beta{\bf r}$, where $m_0$ is the rest mass of the oscillator with angular frequency $\omega$, {\bf{r}} is the position vector and $\beta$ ($\beta=\gamma^0$) is an usual Dirac matrix. Since that was proposed in the literature, several works on the DO have been and continue being performed in different areas of the physics, such as in thermodynamics \citep{Boumali2013,Pacheco}, physics-mathematics \cite{Benitez,Maluf}, nuclear physics \cite{Munarriz,Kulikov,Grineviciute}, quantum chromodynamics \cite{Moshinsky1993,de Lange}, quantum optics \cite{Bermudez,Longhi} and in graphene \cite{Quimbay2013,Boumali,Belouad}. In Refs. \cite{Bakke,Neto}, a DO-type coupling was used to model 2D quantum rings. In 2013, the 1D DO was verified experimentally by J. A. Franco-Villafa\~ne $et$ $al$ \cite{Franco} and recently was studied in the context of the position-dependent mass \cite{Ho} and in the presence of the Aharonov-Bohm-Coulomb system \cite{Oliveira}, topological defects \cite{Salazar,Hosseinpour} and of external electromagnetic fields \cite{O}.

Over the years, the study of noninertial effects generated by rotating frames have also been investigated in the literature \cite{Matsuo2011}, where the best-known effects are the Sagnac \cite{Sagnac,Post}, Barnett \cite{Barnett,Ono}, Einstein-de Hass \cite{Einstein} and Mashhoon \cite{Mashhoon} effects. In the last years, noninertial effects were investigated in some condensed matter systems, such as in the quantum Hall effect \cite{Fischer,Viefers,Matsuo}, Bose-Einstein condensates \cite{Schweikhard,Cooper,Bretin}, fullerene molecules \cite{Shen,Lima}, and in atomic gases \cite{Cooper2008,Lu}. Discussions about the neutrons interferometry induced by Earth's rotation was made in Ref. \cite{Werner}. On the other hand, the study of noninertial effects in relativistic quantum systems also gained relevance and focus of investigations in recent years \cite{Wang2018,Santos,Hosseinpour2015,Castro}. In particular, the DE in a rotating frame has several applications, for instance, is applied in physical problems involving spin currents \cite{Dayi2018,MatsuoPRL}, Sagnac and Hall effects \cite{Anandan,Zubkov}, chiral symmetry \cite{Chernodub}, external magnetic fields \cite{Liu,Chernodub2017}, fullerene molecules \cite{Cavalcante,Gonzalez,Kolesnikov}, nanotubes and carbon nanocones \cite{Cunha,Gomes}, etc.

The present work has as its goal to investigate the influence of noninertial and spin effects on the relativistic and nonrelativistic quantum dynamics of the 2D DO in the magnetic cosmic string background. To model this background, we consider a uniform magnetic field, the Aharonov-Bohm (AB) effect and a deficit angle $\eta$, where $\eta\equiv 1-4G\mu/c^2$ and $\mu>0$ is the linear mass density of a cosmic string (linear gravitational topological defect). According to the literature, the first papers that studied the DO in an inertial frame in the presence of the AB effect with and without the cosmic string spacetime and the spin effects are found in Refs. \cite{Ferkous,Carvalho,Andrade2014}. However, the first papers that studied the DO without magnetic interaction under the influence of noninertial effects with and without the cosmic string spacetime are given in Refs. \cite{Strange,B,Ba}. In particular, our work generalizes the results in Refs. \cite{Ferkous,Andrade2014,Strange} and of other works associated, where we are showing that the dynamics of the system is significantly affected when the noninertial and spin effects are taken into consideration.

This paper is organized as follows. In Sect. \ref{sec2}, we present the cosmic string background as well as the configuration of the external magnetic field and of the AB effect in a uniform rotating frame. In Sect. \ref{sec3}, we investigate the influence of noninertial and spin effects on the quantum dynamics of the 2D DO in the presence of a uniform magnetic field and of the AB effect in the cosmic string background. Next, we determine the two-component Dirac spinor and the energy spectrum for the relativistic bound states. In Sect. \ref{sec4}, we analyze the nonrelativistic limit of our results. Finally, in Sect. \ref{conclusion} we present our conclusions. In this work, we use the natural units ($\hbar=c=G=1$), the cosmic string spacetime with signature $(+,-,-,-)$, and the orthogonal curvilinear coordinates system.

\section{The cosmic string spacetime and the configuration of the magnetic field and of the AB effect in a rotating frame\label{sec2}}

In this section, we configure the curved spacetime background in a rotating frame; whose chosen spacetime is the cosmic string spacetime along of the $Z$-axis, where its line element in general cylindrical coordinates is given by \cite{Ba,Vilenkin}
\ie ds^2_{string}=dT^2-dR^2-\eta^2R^2 d\Phi^2-dZ^2,
\label{lineelement}\fe
being the parameter $\eta$ defined in the range $0<\eta<1$ and the geometry characterized by this line element has a conical singularity that gives rise to the curvature centered on the cosmic string axis ($Z$-axis); however, in all other places, the curvature is null. In particular, this conical singularity is represented by the following curvature tensor
\ie R^{\rho,\varphi}_{\rho,\varphi}=\frac{1-\eta}{4\eta}\delta_2(\bf r),
\label{curvature}\fe
where $\delta_2(\bf r)$ is the 2D Dirac delta. In condensed matter physics, it is already well known that linear topological defects as disclinations and dislocations can be described through a line element in the same way as a topological defect in general relativity \cite{Katanaev,Kleinert}. It is worth mentioning that in the case of the cosmic string, the spatial part of its line element corresponds to the line element of a disclination. In that way, when we take the nonrelativistic limit of the DE, we can extend this formalism to the solid-state physics context \cite{Bakke2010}.

As we are interested in working also in a rotating frame, we must perform the following coordinate transformations
\ie T=t, \ \ R=\rho, \ \ \Phi=\varphi+\Omega t, \ \ Z=z,
\label{coordinate}\fe
where $\Omega=\Omega_z>0$ is the constant angular velocity (not an angular frequency) of the rotating frame, $t\in(-\infty,\infty)$ is the temporal coordinate, $\rho=(x^2+y^2)^{1/2}\in[0,\infty)$ is the radial coordinate, and $\varphi\in[0,2\pi]$ is the angular coordinate (polar angle), respectively. In particular, the coordinates $t$, $\rho$, $\varphi$, and $z$ in the rotating frame (which rotates together with the system) coincide with the corresponding coordinates of the inertial laboratory frame, i.e., $t=t_{lab}$, $\rho=\rho_{lab}$, $\varphi=\varphi_{lab}$, and $z=z_{lab}$ \cite{Chernodub}.

With the transformations in (\ref{coordinate}), the line element (\ref{lineelement}) becomes
\ie ds_{string-rotating}^2=(1-V^2)\left(dt-\frac{V\eta\rho}{(1-V^2)}d\varphi\right)^2-d\rho^2-\frac{\eta^2\rho^2}{(1-V^2)}d\varphi^2-dz^2,
\label{lineelement2}\fe
where $V\equiv\Omega\eta\rho>0$ is the ratio between the velocities of the rotating frame and the of light and satisfies $V<1$ (causality requirement).

Thus, with the line element (\ref{lineelement2}), we need to construct a local reference frame where the observer will be placed (the laboratory frame); consequently, we can define from this the Dirac matrices in the rotating curved spacetime. In this way, a local reference frame can be built through of a noncoordinate basis given by $\hat{\theta}^a=e^a_{\ \mu}(x)dx^\mu$, which its components $e^a_{\ \mu}(x)$ satisfy the following relation \cite{Ba,Bakke2010}
\ie \mathrm{g}_{\mu\nu}(x)=e^a_{\ \mu}(x)e^b_{\ \nu}(x)\eta_{ab},\ \  \eta_{ab}=diag(+1,-1,-1,-1),
\label{tensor}\fe
where $g_{\mu\nu}(x)$ is the curved metric tensor, $\eta_{ab}$ is the Minkowski metric tensor (in Cartesian coordinates) and the indices $\mu,\nu=t,\rho,\varphi,z$ and $a,b=0,1,2,3$ indicate the general and local reference frames, respectively. The components of the noncoordinate basis $e^a_{\ \mu}(x)$ are called $tetrads$ (or $vierbein$), whose inverse is defined as $dx^\mu=e_{\ a}^\mu(x)\hat{\theta}^a$, where $e^{a}_{\ \mu}(x)e^{\mu}_{\ b}(x)=\delta^{a}_{\ b}$ and $e^{\mu}_{\ a}(x)e^{a}_{\ \nu}(x)=\delta^{\mu}_{\ \nu}(x)$ must be satisfied. In addition, we can now rewrite now the line element \eqref{lineelement2} in terms of non coordinate basis as \cite{Strange}
\ie ds_{string-rotating}^2=g_{\mu\nu}(x)dx^\mu dx^\nu=\eta_{ab}\hat{\theta}^a\otimes\hat{\theta}^b=(\hat{\theta}^0)^2-(\hat{\theta}^1)^2-(\hat{\theta}^2)^2-(\hat{\theta}^3)^2,
\label{lineelement3}\fe
where
\ie \hat{\theta}^0=\sqrt{1-V^2}\left(dt-\frac{V\eta\rho}{(1-V^2)}d\varphi\right), \ \ \hat{\theta}^1=d\rho, \ \ \hat{\theta}^2=\frac{\eta\rho}{\sqrt{1-V^2}}d\varphi, \ \ \hat{\theta}^3=dz,
\label{bases}\fe
being the metric tensor given by
\ie g_{\mu\nu}(x)=\left(\begin{array}{cccc}
1-V^2 & \ 0 & -V\eta\rho & \ \ 0	\\
0 & -1 &  0 & \ \ 0	\\
-V\eta\rho & \ 0 & -(\eta\rho)^2 & \ \ 0	\\
0 & \ 0 & 0 & -1
\end{array}\right).
\label{metric1}\fe

As we can see in \eqref{bases}, a given quantity (or parameter) with the indices $a,b$ does not mean that such quantity is in inertial Minkowski spacetime in Cartesian coordinates. In the case of the metric tensor $\eta_{ab}$, such tensor is written in Cartesian coordinates because is where the DE was originally formulated \cite{Greiner}. For the sake of information, making $\eta\to 1$ in \eqref{lineelement3} (absence of the cosmic string), we get the line element in rotating Minkowski spacetime in cylindrical coordinates (and not in Cartesian coordinates) \cite{Strange}. Already for $\eta\to 1$ and $\Omega\to 0$ in \eqref{metric1} with signature $(-,+,+)$ (absence of the rotating cosmic string), we get the metric tensor in inertial Minkowski spacetime in polar coordinates (and not in Cartesian coordinates) \cite{Villalba}.

Our interest here is to build a rotating frame where there is no torque on the system; consequently, the $tetrads$ and its inverse takes the form
\ie e^{a}_{\ \mu}(x)=\left(\begin{array}{cccc}
\sqrt{1-V^2} & 0 & -\frac{V\eta\rho}{\sqrt{1-V^2}} & 0	\\
0 & 1 & 0 & 0	\\
0 & 0 & \frac{\eta\rho}{\sqrt{1-V^2}} & 0	\\
0 & 0 & 0 & 1
\end{array}\right), \ \ e^{\mu}_{\ a}(x)=\left(\begin{array}{cccc}
\frac{1}{\sqrt{1-V^2}} & 0 & \frac{V}{\sqrt{1-V^2}} & 0	\\
0 & 1 & 0 & 0 \\
0 & 0 & \frac{\sqrt{1-V^2}}{\eta\rho} & 0	\\
0 & 0 & 0 & 1
\end{array}\right),
\label{tetrads}\fe

With the information about the choice of the local reference frame, we can obtain the one-form connection $\omega^a_{\ b}=\omega^{\ a}_{\mu\ b}(x)dx^\mu$ through the Maurer-Cartan structure equations. In the absence of the torsion (torque), these equations can be written as
\ie d\hat{\theta}^a+\omega^a_{\ b}\wedge\hat{\theta}^b=0,
\label{Maurer-Cartan}\fe
where the operator $d$ is the exterior derivative and the symbol $\wedge$ means the external product. Therefore, the non-null components of the one-form connection are
\ie \omega^{\ 0}_{t\ 1}(x)=\omega^{\ 1}_{t\ 0}(x)=-\frac{V\Omega\eta}{\sqrt{1-V^2}},
\label{one-form}\fe
\ie \omega^{\ 1}_{t\ 2}(x)=-\omega^{\ 2}_{t\ 1}(x)=-\frac{\Omega\eta}{\sqrt{1-V^2}},
\label{two-form}\fe
\ie \omega^{\ 0}_{\rho\ 2}(x)=\omega^2_{\rho\ 0}(x)=\frac{\Omega\eta}{(1-V^2)},
\label{there-form}\fe
\ie \omega^{\ 0}_{\varphi\ 1}(x)=\omega^{\ 1}_{\varphi\ 0}(x)=-\frac{V\eta}{\sqrt{1-V^2}},
\label{four-form}\fe
\ie \omega^{\ 1}_{\varphi\ 2}(x)=-\omega^{\ 2}_{\varphi\ 1}(x)=-\frac{\eta}{\sqrt{1-V^2}}.
\label{five-form}\fe

Now, we will focus our attention on the configuration of the external magnetic field in the cosmic string background in a rotating frame. Therefore, to insert a magnetic interaction into DE due to a particle with electric charge $q$ ($q<0$) we must introduce into DE a minimal electromagnetic coupling given by: $i\gamma^{\mu}(x)\partial_\mu\to i\gamma^{\mu}(x)(\partial_\mu+iqA_\mu(x))$, where $A_\mu(x)=e^a_{\ \mu}(x)A_a$ is the curved electromagnetic field and $\gamma^{\mu}(x)$ are the curved gamma matrices \cite{Andrade2014,Bakke2010,Greiner}. Explicitly, $A_\mu(x)$ is written in the rotating frame $(\Omega\neq0)$ as follow
\ie A_\mu(x)=(0,0,e^2_{\ \varphi}(x)A_2,0)=\left(0,0,-\frac{\eta\rho}{\sqrt{1-V^2}}A_\varphi,0\right), \ \ (A_t=A_\rho=A_z=0),
\label{four-vector}\fe
where $A_\varphi$ is the angular component of the flat electromagnetic field $A_a=(0,0,A_2,0)=(0,0,-A_\varphi,0)$ written in the inertial frame of the observer $(\Omega=0)$. Here, we consider a uniform external magnetic field generated by a infinite solenoid of radius $\rho$ perpendicular to the polar plane given by ${\bf B}_{ext}=B\hat{\bf z}$ ($B=B_z>0$), where the vector potential for this field is ${\bf A}_1=\frac{1}{2\eta}B\rho\hat{\varphi}$ \cite{Oliveira}, and the vector potential of the AB effect generated by a infinite solenoid of radius $a$ ($a<\rho$) perpendicular to the polar plane given by ${\bf A}_2=\frac{\Phi}{2\pi\eta\rho}\hat{\varphi}$, where $\Phi>0$ is the constant magnetic flux \cite{Carvalho,Andrade2014,Aharonov}. In this ways, we have ${\bf A}={\bf A}_1+{\bf A}_2=A_\varphi(\rho)\hat{\varphi}$, where $A_\varphi(\rho)=\frac{1}{2\eta}B\rho+\frac{\Phi}{2\pi\eta\rho}$ and the parameter $\eta$ arises due to the fact that the two solenoids coincide with the axis of symmetry of the cosmic string (``is a correction factor''). However, for $\eta\to 1$ we have the vector potentials in Minkowski spacetime in cylindrical coordinates.

\section{Relativistic quantum dynamics of the 2D Dirac oscillator in the rotating magnetic cosmic string background \label{sec3}}

In this section, we obtain the relativistic bound-state solutions of the 2D DO interacting with a uniform magnetic field and the AB effect in the rotating cosmic string spacetime. However, we initially need to turn our problem in a 2D system (``conical surface''). In this way, using a symmetry under translations in $z$ (translational invariance) and starting from fact that in the $tetrads$ formalism we are free to choose a representation for the matrices gamma, we can turn such matrices in the $2\times 2$ Pauli matrices \cite{Andrade2014,Andrade}. Besides that, these arguments are valid since our vector potentials are intrinsically 2D (depends only on polar radial coordinate). So, the equation of motion for the covariant 2D DO in a curved spacetime (curvilinear coordinates system) interacting with an external electromagnetic field is written by the following expression \cite{Ba}
\ie i\gamma^{\mu}(x)\left(\nabla_\mu(x)+iqA_{\mu}(x)+m_0\omega\rho\gamma^{0}\delta^{\rho}_\mu\right)\psi_c(t,\rho,\varphi)=m_0\psi_c(t,\rho,\varphi), \ \ (\mu=t,\rho,\varphi),
\label{string1}\fe
where $\gamma^{\mu}(x)=e^\mu_{\ a}(x)\gamma^a$ are the curved gamma matrices and $\gamma^a$ are the flat gamma matrices defined in the inertial Minkowski spacetime, $\nabla_\mu(x)=\partial_\mu+\Gamma_\mu(x)$ is the covariant derivative, being $\Gamma_\mu(x)=-\frac{i}{4}\omega_{\mu ab}(x)\sigma^{ab}$ the spinorial connections and $\omega_{\mu ab}(x)$ the spin connections, and $\psi_c$ is the two-component curvilinear Dirac spinor. The connection between $\psi_c$ and the original Dirac spinor $\Psi_D$ is given by $\psi_c=e^{\frac{i\Sigma_3\varphi}{2}}\Psi_D$, where the unitary operator $e^{\frac{i\Sigma_3\varphi}{2}}$ has the function of to transform (originally in Minkowski spacetime) the curvilinear gamma matrices (``hard to work'') into Cartesian fixed matrices (``easy to work'') \cite{Villalba}. Another fact that corroborates with our statement is that these spinors must satisfy the following periodicity conditions of the system: $\psi_c(\varphi\pm 2\pi)=-\psi_c(\varphi)$ and $\Psi_D(\varphi\pm 2\pi)=\Psi_D(\varphi)$, something that is only consistent with the operator $e^{\frac{i\Sigma_3\varphi}{2}}$ making the connection between both spinors.

With the one-form connections given in
\eqref{one-form}, \eqref{two-form}, \eqref{there-form}, \eqref{four-form} and \eqref{five-form}, we obtain
\ie \Gamma_t(x)=-\frac{1}{2}\frac{V\Omega\eta}{\sqrt{1-V^2}}\gamma^0\gamma^1-\frac{i}{2}\frac{\Omega\eta}{\sqrt{1-V^2}}\Sigma^3, \ \ (\Sigma^3=\Sigma_3=i\gamma^1\gamma^2)
\label{spinorial1},\fe
\ie \Gamma_\rho(x)=\frac{1}{2}\frac{\Omega\eta}{(1-V^2)}\gamma^0\gamma^2
\label{spinorial2},\fe
\ie \Gamma_\varphi(x)=-\frac{1}{2}\frac{V\eta}{\sqrt{1-V^2}}\gamma^0\gamma^1-\frac{i}{2}\frac{\eta}{\sqrt{1-V^2}}\Sigma^3
\label{spinorial3},\fe
where implies that
\ie \gamma^\mu(x)\Gamma_\mu(x)=\frac{1}{2}\frac{i\eta\gamma^0\Omega}{(1-V^2)}\Sigma^3+\frac{1}{2\rho}\gamma^1
\label{spinorial}.\fe

In addition, the curved gamma matrices are given by
\ie \gamma^t(x)=\frac{1}{\sqrt{1-V^2}}\gamma^0+\frac{V}{\sqrt{1-V^2}}\gamma^2
\label{gama0},\fe
\ie \gamma^\rho(x)=\gamma^1
\label{gama1},\fe
\ie \gamma^\varphi(x)=\frac{\sqrt{1-V^2}}{\eta\rho}\gamma^2
\label{gama2}.\fe

Thus, using the informations here presented and of the section II, Eq. \eqref{string1} becomes
\begin{eqnarray}\nonumber
& & \frac{i}{\sqrt{1-V^2}}\left(\gamma^0+V\gamma^2\right)\frac{\partial\psi_c}{\partial t}+i\gamma^1\left(\frac{\partial}{\partial\rho}+\frac{1}{2\rho}+m_0\omega\rho\gamma^{0}\right)\psi_c-\frac{\eta\gamma^0}{(1-V^2)}{\bf S}\cdot{\bf\Omega}\psi_c \\
& & +i\gamma^2\left[\frac{1}{\eta\rho}\left(\sqrt{1-V^2}\frac{\partial}{\partial\varphi}+i\frac{\Phi}{\Phi_0}\right)+\frac{ie}{2\eta}B\rho\right]\psi_c=m_0\psi_c,
\label{string2}
\end{eqnarray}
where the term ${\bf S}\cdot{\bf\Omega}$ is called of spin-rotation coupling \cite{Mashhoon,Hehl}, being ${\bf\Omega}=\Omega\hat{\bf z}$, ${\bf S}=\frac{1}{2}{\bf\Sigma}$ is the spin operator, $\Phi_0\equiv\frac{2\pi}{e}$ is the magnetic flux quantum and we assume $q=-e$ ($e>0$). It is also important to mention that this coupling is the origin of a term in the energy spectrum called (rotating) zero-point energy (analogous to the case of the nonrelativistic quantum harmonic oscillator) \cite{Strange}. Moreover, Eq. (\ref{string2}) is in accordance with the curvilinear DO found in Ref. \cite{Villalba}, i.e., the DO in Minkowski spacetime in polar coordinates. Therefore, the spinor in (\ref{string2}) it’s really a curvilinear spinor.  

On the other hand, we see that it is difficult to proceed without a simplification of Eq. (\ref{string2}). To solve analytically this equation, we consider that the linear velocity of the rotating frame being small compared with the velocity of the light ($V^2\ll 1$) \cite{Strange}. Using this condition, Eq. \eqref{string2} becomes
\begin{eqnarray}\nonumber
& & i\left\{\gamma^0\frac{\partial}{\partial t}+\gamma^1\left(\frac{\partial}{\partial\rho}+\frac{1}{2\rho}+m_0\omega\rho\gamma^{0}\right)+\gamma^2\left[\frac{1}{\eta\rho}\left(\frac{\partial}{\partial\varphi}+i\frac{\Phi}{\Phi_0}\right)+\frac{ie}{2\eta}B\rho+\Omega\eta\rho\frac{\partial}{\partial t}\right]\right\}\psi_c \\
& & -(\eta\gamma^0{\bf S}\cdot{\bf\Omega}+m_0)\psi_c=0.
\label{string3}
\end{eqnarray}

Since we are working planar spacetime (2D spacetime), it is convenient to write the gamma matrices $\gamma^a=(\gamma^0,\gamma^1,\gamma^2)=(\gamma_0,-\gamma_1,-\gamma_2)$ and the matrix $\Sigma^3$ in terms of the $2\times 2$ Pauli matrices, i.e., $\gamma_1=\sigma_3\sigma_1=i\sigma_2$, $\gamma_2=s\sigma_3\sigma_2=-is\sigma_1$ and $\gamma^0=\Sigma^3=\sigma_3$ \cite{Andrade2014,Andrade,Villalba,Greiner}. In particular, the parameter $s$ (spin parameter) characterizes the two spin states of the particle, with $s=+1$ for spin ``up'' ($\uparrow$) and $s=-1$ for spin ``down'' ($\downarrow$), respectively. By using this information and the Pauli matrices in the form
\ie \sigma_1=\left(
    \begin{array}{cc}
      0\ & \ 1  \\
      1\ & \ 0 \\
    \end{array}
  \right), \ \  \sigma_2=\left(
    \begin{array}{cc}
      0\ & -i  \\
      i\ & \ 0 \\
    \end{array}
  \right), \ \  \sigma_3=\left(
    \begin{array}{cc}
      1\ & \ 0  \\
      0\ & -1 \\
    \end{array}
  \right),
\label{matrices} \fe
and defining the following ansatz for the two-component spinor \cite{Villalba,Rubens}
\ie \psi_c(t,\rho,\varphi)=\frac{e^{i(m_j\varphi-Et)}}{\sqrt{2\pi}}\left(
           \begin{array}{c}
            R_{+}(\rho) \\
            iR_{-}(\rho) \\
           \end{array}
         \right), \ \ (m_j=\pm1/2,\pm3/2,\pm5/2,\ldots)
\label{spinor},\fe
we obtain from \eqref{string3} a system of two first-order coupled differential equations given by
\ie\left(m_0+\frac{\eta\Omega}{2}-E\right)R_+(\rho)=\left[\frac{d}{d\rho}-m_0\bar{\Omega}\rho+\frac{s}{\eta\rho}\left(m_j+\frac{\Phi}{\Phi_0}+\frac{s\eta}{2}\right)\right]R_-(\rho)
\label{string4}, \fe
\ie\left(m_0+\frac{\eta\Omega}{2}+E\right)R_-(\rho)=\left[\frac{d}{d\rho}+m_0\bar{\Omega}\rho-\frac{s}{\eta\rho}\left(m_j+\frac{\Phi}{\Phi_0}-\frac{s\eta}{2}\right)\right]R_+(\rho)
\label{string5},\fe
where $\bar{\Omega}\equiv(\omega-s\frac{\omega_c}{2\eta}+s\frac{\eta\Omega E}{m_0})$ is an effective angular frequency and $\omega_c=\frac{eB}{m_0}>0$ is the cyclotron frequency (angular velocity) of the particle in the plane, $E$ is the relativistic total energy, $\frac{1}{\sqrt{2\pi}}$ is a factor that arises from normalization of the angular part of the spinor and $m_j$ is the curvilinear total magnetic quantum number (arises from condition $\psi_c(\varphi\pm 2\pi)=-\psi_c(\varphi)$). Here, the connection between $m_j$ and $m_l$, where $m_l$ is the curvilinear orbital magnetic quantum number, is given as follows
\ie J_z\psi_c(t,\rho,\varphi)=-i\frac{\partial\psi_c(t,\rho,\varphi)}{\partial\varphi}=m_j\psi_c(t,\rho,\varphi)=(m_l+m_s)\psi_c(t,\rho,\varphi)
\label{J},\fe
where $J_z=L_z+S_z$ is the $z$-component of the total angular momentum ${\bf J}$, being $S_z=\frac{1}{2}\sigma_z$, $m_s=\pm\frac{1}{2}$ is the curvilinear spin magnetic quantum number (spin up or spin down) and the values of $m_l$ are given by $m_l=0,\pm 1,\pm 2,\ldots$ \cite{Villalba}.

Substituting now \eqref{string5} into \eqref{string4} and vice versa, we get two differential equations written compactly as
\ie \left[\frac{d^2}{d\rho^2}+\frac{1}{\rho}\frac{d}{d\rho}-\frac{\gamma_{r}^2}{\eta^2\rho^2}-m^2\bar{\Omega}^2\rho^2+E_{r}\right]R_r(\rho)=0, \ \ (r=\pm 1)
\label{string6}, \fe
where we defined
\ie \gamma_{r}\equiv m_j+\frac{\Phi}{\Phi_0}-\frac{rs\eta}{2}, \ \  E_{r}\equiv E^2-\left(m_0+\frac{\eta\Omega}{2}\right)^2+\frac{2sm_0\bar{\Omega}}{\eta}\gamma_{r}+2rm_0\bar{\Omega}
\label{string7}, \fe
being $R_r(\rho)$ real radial functions and $r$ characterizes the two components of the spinor, being that $r=+1$ describes a particle with spin up ($s=+1$) or down ($s=-1$) and $r=-1$ describes an antiparticle with spin up ($s=+1$) or down ($s=-1$), respectively.

In order to solve Eq. \eqref{string6}, we will introduce a new dimensionless variable given by $\tau=m_0\bar{\Omega}\rho^2$ ($\bar{\Omega}>0$). Thereby, by making a variable change, Eq. \eqref{string6} becomes
\ie \left[\tau\frac{d^{2}}{d\tau^{2}}+\frac{d}{d\tau}-\frac{\gamma_{r}^{2}}{4\eta^2\tau}-\frac{\tau}{4}+\mathcal{E}_{r}\right]R_r(\tau)=0,
\label{string8}\fe 
where
\ie \mathcal{E}_{r}\equiv\frac{E_{r}}{4m_0\bar{\Omega}}.
\label{E}\fe

Analyzing the asymptotic behavior of Eq. \eqref{string8} for $\tau\to 0$  and $\tau\to\infty$, we can write a regular solution for this equation as
\ie R_r(\tau)=C_r\tau^{\frac{\vert\gamma_{r}\vert}{2\eta}}e^{-\frac{\tau}{2}}F_r(\tau)
\label{string9},\fe 
where $C_r>0$ are normalization constants, $F_r(\tau)$ are unknown functions to be determined and $R_r(\tau)$ must satisfy the following boundary conditions to be a physically acceptable solution (normalizable solution)
\ie R_r(\tau\to 0)=R_r(\tau\to\infty )=0
\label{conditions}.\fe 

In this way, substituting \eqref{string9} into Eq. \eqref{string8}, we obtain
\ie \left[\tau\frac{d^{2}}{d\tau^{2}}+(\vert\bar{\gamma}_r\vert-\tau)\frac{d}{d\tau}-\left(\frac{\vert\bar{\gamma}_r\vert}{2}-\mathcal{E}_{r}\right)\right]F_r(\tau)=0,
\label{string10}
\end{equation}
where
\ie \vert\bar{\gamma}_r\vert\equiv\frac{\vert\gamma_r\vert}{\eta}+1.
\label{G}
\end{equation}

It is not difficult to note that Eq. \eqref{string10} is a generalized Laguerre equation, whose solution are the generalized Laguerre polynomials and is denoted by \cite{Andrade,Abramowitz}
\ie F_r(\tau)=L^{\frac{\vert\gamma_r\vert}{\eta}}_n (\tau).
\label{hypergeometric}
\end{equation}

However, for that the Dirac spinor becomes a solution finite (normalizable), is necessary that the generalized Laguerre polynomials be a polynomial of degree $n$, consequently, the parameter $\frac{\vert\bar{\gamma}_r\vert}{2}-\mathcal{E}_r$ should be a non-positive integer, i.e., $\frac{\vert\bar{\gamma}_r\vert}{2}-\mathcal{E}_r=-n$, where $n=0,1,2,\ldots$ (quantum number). Therefore, we obtain from this condition (quantization condition) the following energy spectrum of the DO under the influence of noninertial and spin effects in the magnetic cosmic string spacetime
\ie E^\sigma_{n,m_j,r,s}=2s\eta\Omega N_r+\sigma\sqrt{(2\eta\Omega N_r)^2+(m_0+\eta E_0)^2+4m_0\vert\omega_s\vert N_r},
\label{spectrum}
\end{equation}
where
\ie N_r\equiv\left(n+\frac{1-r}{2}+\frac{\Gamma_r}{2\eta}\right), \ \ E_0\equiv\frac{\Omega}{2},
\label{N}
\end{equation}
being $N_r$ an effective quantum number, $\sigma= 1$ corresponds to the positive energy states (particle or DO), $\sigma=-1$ corresponds to the negative energy states (antiparticle or anti-DO), and the quantities $\Gamma_r$ and $\omega_s$ are given by $\Gamma_r\equiv\vert\gamma_r\vert-s\gamma_r$ and $\omega_s\equiv(\omega-s\frac{\omega_c}{2\eta})$, respectively. We see that the energy spectrum \eqref{spectrum} explicitly depends on the spin parameter $s$, magnetic flux $\Phi$ of the AB effect, cyclotron frequency $\omega_c$ generated by the external magnetic field, zero-point energy $E_0$, angular velocity $\Omega$ of the rotating frame, and of the deficit angle $\eta$ associated to topology of the cosmic string. In particular, this spectrum satisfies: $E^\sigma_{n,m_j,r,s}\left(\Phi\pm\Phi_0\right)=E^\sigma_{n,m_j\pm 1,r,s}(\Phi)$, therefore, the spectrum is a periodic function with periodicity $\pm\Phi_0$ \cite{Vitoria}. We note that the presence of $\Omega$ causes an asymmetry in the spectrum, since for $\Omega\to 0$ we have a symmetrical spectrum \cite{Strange}. This asymmetry comes from the fact that in some cases we have $\vert E^+_{n,m_j,r,s}\vert>\vert E^-_{n,m_j,r,s}\vert$ or $\vert E^+_{n,m_j,r,s}\vert<\vert E^-_{n,m_j,r,s}\vert$ depending on the values of $\sigma$ and $s$. In Table I, are shows the four settings of the relativistic bound state energies for the possible combinations of the parameters $\sigma$ and $s$.
\begin{center}\label{t}
\begin{table*}[h]
\centering
\caption{Relativistic bound state energies for the possible combinations of $\sigma$ and $s$.}
\def\arraystretch{1.1}
\begin{tabular}{cccc}
\hline
Setting & $E^\sigma_{n,m_j,r,s}$ & $\sigma$ & $s$ \\ \hline
1 & $E^+_{n,m_j,+,+}=2\eta\Omega N_++\sqrt{(2\eta\Omega N_+)^2+(m_0+\eta E_0)^2+4m_0\vert\omega_+\vert N_+}$       & + 1      & \ + 1 \\
2 & $E^-_{n,m_j,-,-}=-2\eta\Omega N_--\sqrt{(2\eta\Omega N_-)^2+(m_0+\eta E_0)^2+4m_0\vert\omega_-\vert N_-}$       & - 1      & \ \ - 1 \\
3 & $E^+_{n,m_j,+,-}=-2\eta\Omega N_++\sqrt{(2\eta\Omega N_+)^2+(m_0+\eta E_0)^2+4m_0\vert\omega_-\vert N_+}$       & + 1      & \ \ - 1 \\
4 & $E^-_{n,m_j,-,+}=2\eta\Omega N_--\sqrt{(2\eta\Omega N_-)^2+(m_0+\eta E_0)^2+4m_0\vert\omega_+\vert N_-}$       & - 1      & + 1 \\ \hline
\end{tabular}
\end{table*}
\end{center}

According to table I, the settings $1$ and $3$ describe the spectrum of a particle ($\sigma=r=+1$) with spin up or down, where its energies are larger for the case of the spin up ($E^+_{n,m_j,+,+}>E^+_{n,m_j,+,-}$), while the settings $2$ and $4$ describe the spectrum of an antiparticle ($\sigma=r=-1$) with spin down or up, where its energies (in absolute values) are larger for the case of the spin down ($\vert E^-_{n,m_j,-,-}\vert>\vert E^-_{n,m_j,-,+}\vert$), respectively. Besides that, based on the fact that $N_->N_+$ and $\omega_->\omega_+$, we verified that the energies of the antiparticle with spin down are larger than those of the particle with spin up or down, i.e., $\vert E^-_{n,m_j,-,-}\vert>\vert E^+_{n,m_j,+,\pm}\vert$ (asymmetry in the spectrum). Still according to table I, we see that the quantities $\Omega$, $\eta$ and $\omega_c$ and the quantum number $N_r$ have the function of increase the values of the spectrum, for instance, in the limit $\eta\to 0$ (extremely dense cosmic string) or $N_r=\omega_c=\Omega\to\infty$, we have $\vert E^\sigma_{n,m_j,r,s}\vert\to\infty$. However, the quantity $\Phi$ has the function of increasing or decreasing the values of the spectrum depending on the values of $s$, for instance, for $\Phi\to\infty$ and $s=-1$ we have $\vert E^\sigma_{n,m_j,r,s}\vert\to\infty$ ($N_r\to\infty$), while for $\Phi\to\infty$ and $s=+1$ we have $\vert E^\sigma_{n,m_j,r,s}\vert\to\sigma(m_0+\eta E_0)$ ($N_r\to 0$). Last but not least, we see that even in the absence of the AB effect ($\Phi=0$), string cosmic background ($\eta=1$), uniform magnetic field and of the DO ($\omega_s=0$), the particle and antiparticle still have a discrete energy spectrum, in this case, we can say that the rotating frame (or centripetal force) quantizes its energies.

Now, comparing the spectrum \eqref{spectrum} with the literature, we verified that in the absence of the noninertial effects ($\Omega\to 0$), of the AB effect ($\Phi\to 0$) and of the cosmic string spacetime ($\eta\to 1$), with $r=+1$, we recover the usual spectrum of the DO in a flat inertial frame for $\omega_c\to 0$ with $s=\pm 1$ and $\gamma_r=m_j-\frac{s}{2}\to m_l$ \cite{Andrade}, for $\omega_c\to -\omega_c\neq0$ (charge conjugation) with $s=+1$ \cite{Villalba}, and for $\omega_c\neq 0$ with $s=+1$ and $m_j>0$ ($\gamma_r=0$) \cite{Mandal,Quimbay}. Already in the limits $\Phi=\omega_c\to 0$ and $\eta\to 1$ with $r=s=+1$ and $m_j>0$, we recover the usual spectrum of the DO in a flat rotating frame \cite{Strange}. Now, in the limits $\Omega=\omega_c\to 0$ with $r=+1$, we recover the usual spectrum of the DO under the influence of the AB effect and spin effects in an inertial frame for $\eta\to 1$ \cite{Ferkous} and for $\eta\neq 1$ \cite{Andrade2014}. On the other hand, in the limits $\Omega=\omega=\Phi\to0$ and $\eta\to 1$ with $m_j>0$, $s=+1$ and $r=-1$, we recover the relativistic Landau levels for a planar Dirac particle in a flat inertial frame \cite{Lamata,Schakel,Miransky}. From the above, we see clearly that our relativistic spectrum generalizes some particular cases of the literature when $\Omega$, $\omega_c$, $\omega$, $\Phi$, $s$ or $\eta$ are excluded from the system.

From here on let us concentrate our attention on the form of the original Dirac spinor for the relativistic bound states. Substituting the variable $\tau=m_0\bar{\Omega}\rho^2$ in the radial functions \eqref{string9}, the two-component curvilinear spinor \eqref{spinor} takes the form
\ie \psi_c(t,\rho,\varphi)=\frac{e^{i(m_j\varphi-Et)}}{\sqrt{2\pi}}\left(
           \begin{array}{c}
            D_+\rho^{\frac{\vert\gamma_{+}\vert}{\eta}}e^{-\frac{m_0\bar{\Omega}\rho^2}{2}}L^{\frac{\vert\gamma_+\vert}{\eta}}_n (m_0\bar{\Omega}\rho^2) \\
             iD_-\rho^{\frac{\vert\gamma_{-}\vert}{\eta}}e^{-\frac{m_0\bar{\Omega}\rho^2}{2}}L^{\frac{\vert\gamma_-\vert}{\eta}}_n (m_0\bar{\Omega}\rho^2)  \\
           \end{array}
         \right),
\label{spinor2} 
\end{equation}
where
\ie D_r\equiv C_r(m_0\bar{\Omega})^{\frac{\vert\gamma_r\vert}{2\eta}}>0, \ \ (r=\pm 1).
\label{C}
\end{equation}

Thus, based on fact that $\psi_c=e^{\frac{i\Sigma_3\varphi}{2}}\Psi_D$, being $\Sigma_3=\sigma_3$=diag$(+1,-1)$, the original Dirac spinor is written as follows
\ie \Psi_D(t,\rho,\varphi)=\frac{e^{i(m_l\varphi-Et)}}{\sqrt{2\pi}}\left(
           \begin{array}{c}
            D_+\rho^{\frac{\vert\gamma_{+}\vert}{\eta}}e^{-\frac{m_0\bar{\Omega}\rho^2}{2}}L^{\frac{\vert\gamma_+\vert}{\eta}}_n (m_0\bar{\Omega}\rho^2)  \\
             iD_-\rho^{\frac{\vert\gamma_{-}\vert}{\eta}}e^{-\frac{m_0\bar{\Omega}\rho^2}{2}}L^{\frac{\vert\gamma_-\vert}{\eta}}_n (m_0\bar{\Omega}\rho^2)  \\
           \end{array}
         \right),
\label{spinor3} 
\end{equation}
where we use by simplicity the relation $m_l=m_j\mp\frac{1}{2}$.

It should be noted that our Dirac spinor simultaneously incorporates the positive and negative values of the quantum number $m_l$ (or $m_j$), which does not happen, for instance, in Ref. \cite{Villalba}. However, the temporal and angular parts of the spinor \eqref{spinor3} are equal to the spinor of Ref. \cite{Villalba}, as expected. From the practical point of view, an of the advantages of we have a spinor with the setting \eqref{spinor3} is the possibility of calculating the physical observables more faster and direct than if we had two spinors, one for each value of $m_l$. Also, taking the limits $\omega=\omega_c=\Phi\to 0$ and $\eta\to 1$, we get the spinor in a rotating frame under the influence of spin effects (the spinor still is normalizable).

\section{Nonrelativistic limit\label{sec4}}

In this section, we analyze the nonrelativistic limit (low energy regime) of our results. To get this limit is necessary to consider that most of the total energy of the system stays concentrated in the rest energy of the particle, i.e., $E\cong m_0+\varepsilon$, where $m_0\gg\varepsilon$ and $m_0\gg\Omega$. So, applying this prescription in Eq. \eqref{string6}, we obtain the following Schr\"{o}dinger-Pauli oscillator (SPO) under the influence of noninertial and spin effects in the magnetic cosmic string spacetime (or in the magnetic conical Euclidean space)
\ie\left[H_{QHO-type}+H_{Zeeman-type}-\left(\frac{2\omega}{\eta}{\bf S}+{\bf\Omega}\right)\cdot{\bf L}-r\frac{\bar{\Omega}}{\eta}\frac{\Phi}{\Phi_0}+\eta E_0\right]\Psi_P(t,\rho,\varphi)=i\frac{\partial\Psi_P(t,\rho,\varphi)}{\partial t},
\label{limit1}\fe
where
\ie H_{QHO-type}=-\frac{1}{2m_0}\left(\frac{\partial^{2}}{\partial\rho^{2}}+\frac{1}{\rho}\frac{\partial}{\partial\rho}-\frac{\ell^2_z}{\eta^2\rho^{2}}\right)+\frac{1}{2}m_0\bar{\Omega}^{2}\rho^{2}, \ \ H_{Zeeman-type}=\frac{1}{2\eta^2}\boldsymbol{\omega_c}\cdot{\bf L},
\label{H}\fe
with
\ie \ell_z=L_z+\frac{\Phi}{\Phi_0}, \ \ \bar{\Omega}=\omega-\frac{r}{\eta}\omega_c+2\eta{\bf S}\cdot{\bf\Omega}, \ \ {\bf\Omega}=\Omega\hat{\bf z},\ \ {\bf S}=\frac{1}{2}\boldsymbol{\sigma}, \ \ L_z=-i\frac{\partial}{\partial\varphi},
\label{limit2}\fe
being $H_{QHO-type}$ the QHO-type Hamiltonian (explaining why this system is called DO), $H_{Zeeman-type}$ is the Zeeman-type Hamiltonian, $E_0=\frac{\Omega}{2}$ is the zero-point energy, and $\Psi_P(t,\rho,\varphi)=\frac{e^{i(\bar{m}_l\varphi-\varepsilon t)}}{\sqrt{2\pi}}(R_+(\rho),R_-(\rho))^T$ ($\bar{m}_l\equiv m_j\mp\frac{\eta}{2}$) is the Pauli spinor and satisfies $S_z\Psi_P=\frac{s}{2}\Psi_P$. Here, we have $r\equiv s=\pm 1$ and $rs=1$, where $s=+1$ and $s=-1$ describes a particle (or SPO) with spin up or down, respectively. We verify that the term ${\bf S\cdot L}$ describes a spin-orbit coupling of strength $\omega/\hbar$ (restoring the factor $\hbar$), while the term ${\bf S}\cdot\boldsymbol{\Omega}$ describe the spin-rotation coupling. Also, we verified that in the limits $\Omega=\omega_c\to 0$ and $\eta\to 1$, Eq. (\ref{limit1}) is reduced to the QHO in a flat inertial frame under the influence of spin effects for $\Phi\to 0$ \cite{Andrade} and for $\Phi\neq 0$ \cite{Ferkous}.

Now, using the prescription $E\cong\varepsilon+m_0$ in \eqref{spectrum} (with $m_0\gg\varepsilon$ and $m_0\gg\Omega$), we obtain the following nonrelativistic energy spectrum of the SPO under the influence of noninertial and spin effects in the magnetic conical Euclidean space
\ie\varepsilon_{n,\bar{m}_l,s}=\eta E_0+\left(2n+1-s+\frac{\vert\alpha\vert-s\alpha}{\eta}\right)(s\eta\Omega+\vert\omega_s\vert)>0,  \ \  \alpha\equiv\bar{m}_l+\frac{\Phi}{\Phi_0}, \ \ \bar{m}_l=m_j-\frac{\eta}{2}.
\label{spectrum2}
\end{equation}

We see that besides of the spectrum \eqref{spectrum2} to depend on the spin parameter $s$, magnetic flux $\Phi$, cyclotron frequency $\omega_c$, zero-point energy $E_0$, angular velocity $\Omega$, and of the deficit angle $\eta$ associated to topology of a conical space, is a periodic function with periodicity $\pm\Phi_0$ \cite{Vitoria}. In addition, this spectrum always increases with the increasing of the quantum number $n$ ($\varepsilon_{n\to\infty}\to\infty$), however, increases or non depending on the values of $\bar{m}_l$ and $s$, for instance, for $\bar{m}_l>0$ and $s=+1$, we have $\varepsilon_{\bar{m}_l\to\infty}\to\eta E_0+2n(\eta\Omega+\vert\omega_+\vert)$, while for $\bar{m}_l>0$ and $s=-1$, we have $\varepsilon_{\bar{m}_l\to\infty}\to\infty$ (with $\vert\omega_-\vert>\eta\Omega$). On the other hand, for $\bar{m}_l<0$ and $s=+1$, we have $\varepsilon_{\bar{m}_l\to-\infty}\to\infty$, while for $\bar{m}_l<0$ and $s=-1$, we have $\varepsilon_{\bar{m}_l\to-\infty}\to\eta E_0+2(n+1)(-\eta\Omega+\vert\omega_-\vert)$. Similar to the relativistic case, we see that the parameters $\eta$ and $\Omega$ have the function of increase the values of the spectrum, i.e., in the limit $\eta\to 0$ (extremely conical space) or $\Omega\to\infty$, we have $\varepsilon_{n,\bar{m}_l,s}\to\infty$ (with $\vert\omega_s\vert>\eta\Omega$).

So, comparing the nonrelativistic spectrum \eqref{spectrum2} with the literature, we verified that in the limits $\omega_c=\Phi\to 0$ and $\eta\to 1$ with $\bar{m}_l=m_l\geq 0$ and $s=+1$, we recover the spectrum of the QHO in a flat rotating frame \cite{Strange}. Already in the limits $\Omega=\omega_c\to 0$ and $\eta\to 1$, we recover the usual spectrum of the QHO in a flat inertial frame under the influence of spin effects for $\Phi\to 0$ \cite{Andrade} and for $\Phi\neq 0$ \cite{Ferkous}. On the other hand, in the limits $\Omega=\Phi\to 0$ and $\eta\to 1$ with $m_l\geq 0$, we recover the usual spectrum of the QHO under the influence of a uniform magnetic field and spin effects in a flat inertial frame \cite{Quimbay}. Finally, in the limit $\Omega=\omega_c\to 0$, we recover the spectrum of the QHO in the cosmic string background under the influence of spin effects \cite{Andrade2014}. From the above, we see categorically that our nonrelativistic spectrum generalizes some nonrelativistic particular cases of the literature when $\Omega$, $\omega_c$, $\Phi$, $s$ or $\eta$ are excluded from the system.

\section{Conclusion\label{conclusion}}

In this paper, we study the influence of noninertial and spin effects on the relativistic and nonrelativistic quantum dynamics of the 2D DO in the magnetic cosmic string background. To model this background, we consider an external uniform magnetic field, the AB effect and a deficit angle $\eta$ of a cosmic string. Posteriorly, we adopted the polar coordinate system and we analyze the asymptotic behavior of our resulting differential equation, where we obtain as results a generalized Laguerre equation. From this result, we determine the bound-state solutions of the system, given by the two-component Dirac spinor and the relativistic energy spectrum. We verified that this spinor is written in terms of the generalized Laguerre polynomials and this spectrum explicitly depends on the quantum numbers $n$ and $m_j$, angular velocity $\Omega$ and parameter $s$ associated to the noninertial and spin effects, magnetic flux $\Phi$ of the AB effect, cyclotron frequency $\omega_c$ generated by the magnetic field, zero-point energy $E_0$, and of the deficit angle $\eta$ associated to topology of the cosmic string. 

In particular, we note that besides of the relativistic spectrum to be a periodic function with periodicity $\pm\Phi_0$, where $\Phi_0$ is the magnetic flux quantum, and asymmetric due to presence of $\Omega$, its values infinitely increases when $\eta\to 0$ (extremely dense cosmic string) or $n=m_j=\omega_c=\Omega\to\infty$. However, the quantity $\Phi$ has the function of increasing or decreasing the values of the spectrum depending on the values of the spin parameter $s$ and of the quantum $m_j$. We also note that the spectrum of the particle or DO (antiparticle or anti-DO) are larger for the case of the spin up (down). In addition, due to asymmetry in the spectrum, we verified that the energies of the antiparticle with spin down are larger than of the particle with spin up or down. Now, comparing our relativistic spectrum with other works, we verified that this spectrum generalizes some particular cases of the literature when $\Omega$, $\omega_c$, $\Phi$, $s$ or $\eta$ are excluded from the system.

Finally, we study the nonrelativistic limit of our results. For instance, considering that most of the total energy of the system stays concentrated in the rest energy of the particle, we obtain the 2D SPO under the influence of noninertial and spin effects in the magnetic cosmic string spacetime (or magnetic conical Euclidean space). We see that this oscillator is written in terms of two Hamiltonians: one QHO-type and other Zeeman-type, and also depends of two types of couplings: the spin-orbit coupling, given by ${\bf S\cdot L}$, and the spin-rotation coupling, given by ${\bf S}\cdot\boldsymbol{\Omega}$. Besides, we see that in the limits $\Omega=\omega_c\to 0$ and $\eta\to 1$, we get the QHO in a flat inertial frame under the influence of spin effects for $\Phi\to 0$ and $\Phi\neq 0$. With respect to the nonrelativistic energy spectrum, such spectrum has some similarities with the relativistic case, i.e., depend on $n$, $m_j$, $s$, $\Phi$, $\omega_c$, $E_0$, $\Omega$, and $\eta$, is a periodic function with periodicity $\pm\Phi_0$ and increase in values as the increase of $n$, $\eta$ and $\Omega$, however, increases or non depending on the values of $m_j$ and $s$. Thus, comparing our nonrelativistic spectrum with other work, we verified that this spectrum generalizes some particular cases of the literature when $\Omega$, $\omega_c$, $\Phi$, $s$ or $\eta$ are excluded from the system.

\section*{Acknowledgments}
\hspace{0.5cm} The author would like to thank the Conselho Nacional de Desenvolvimento Cient\'{\i}fico e Tecnol\'{o}gico (CNPq) for financial support.


\begin{thebibliography}{99}

\bibitem{Moshinsky} Moshinsky, M., Szczepaniak, A.: J. Phys. A: Math. Gen. {\bf 22}, L817 (1989)

\bibitem{Boumali2013} Boumali, A., Hassanabadi, H.: Eur. Phys. J. Plus {\bf 128}, 124 (2013)

\bibitem{Pacheco} Pacheco, M. H., Maluf, R. V., Almeida, C. A. S., Landim, R. R.: Europhys. Lett. {\bf 108}, 10005 (2014)

\bibitem{Benitez} Ben\'itez, J., Martinez-y-Romero, R. P., N\'unez-Y\'epez, H. N., Salas-Brito, A. L.: Phys. Rev. Lett. {\bf 64}, 1643 (1990)

\bibitem{Maluf} Maluf, R. V.: Int. J. Mod. Phys. A {\bf 26}, 4991 (2011)

\bibitem{Munarriz} Munarriz, J., Dominguez-Adame, F., Lima, R. P. A.: Phys. Lett. A {\bf 376}, 3475 (2012)

\bibitem{Kulikov} Kulikov, D. A., Uvarov, I. V., Yaroshenko, A. P.: Cent. Eur. J. Phys. {\bf 11}, 1006 (2013)

\bibitem{Grineviciute} Grineviciute, J., Halderson, D.: Phys. Rev. C {\bf 85}, 054617 (2012)

\bibitem{Moshinsky1993} Moshinsky, M., Loyola, G.: Found. Phys. {\bf 23}, 197 (1993)

\bibitem{de Lange} de Lange, O. L.: J. Math. Phys. {\bf 32}, 1296 (1991)

\bibitem{Bermudez} Bermudez, A., Martin-Delgado, M. A., Solano, E.: Phys. Rev. A {\bf 76}, 041801 (2007)

\bibitem{Longhi} Longhi, S.: Opt. Lett. {\bf 35}, 1302 (2010)

\bibitem{Quimbay2013} Quimbay, C., Strange, P.: Graphene physics via the Dirac oscillator in (2+1) dimensions, arXiv:1311.2021, (2013) [cond-mat.mes-hall].

\bibitem{Boumali} Boumali, A.: Phys. Scrip. {\bf 90}, 045702 (2015)

\bibitem{Belouad} Belouad, A., Jellal, A., Zahidi, Y.: Phys. Lett. A {\bf 380}, 773 (2016)

\bibitem{Bakke} Bakke, K., Furtado, C.: Phys. Lett. A {\bf 376}, 1269 (2012)

\bibitem{Neto} Neto, J. A., Bueno, M. J., Furtado, C.: Ann. Phys. {\bf 373}, 273 (2016)

\bibitem{Franco} Franco-Villafa\~ne, J. A., Sadurn\'i, E., Barkhofen, S., Kuhl, U., Mortessagne, F., Seligman, T. H.: Phys. Rev. Lett. {\bf 111}, 170405 (2013)

\bibitem{Ho} Ho, C. L., Roy, P.: Europhys. Lett. {\bf 124}, 60003 (2019)

\bibitem{Oliveira} Oliveira, R. R. S., Maluf, R. V., Almeida, C. A. S.: Ann. Phys. {\bf 400}, 1 (2019)

\bibitem{Salazar} Salazar-Ram\'irez, M.,  Ojeda-Guill\'en, D., Morales-Gonz\'alez, A., Garc\'ia-Ortega, V. H.: Eur. Phys. J. Plus {\bf 134}, 8 (2019)

\bibitem{Hosseinpour} Hosseinpour, M., Hassanabadi, H., de Montigny, M.: Eur. Phys. J. C {\bf 79}, 311 (2019)

\bibitem{O} Oliveira, R. R. S., Maluf, R. V., Almeida, C. A. S., Exact solutions of the Dirac oscillator under the influence of the Aharonov-Casher effect in the cosmic string background, arXiv:1810.11149, (2018) [quant-ph]

\bibitem{Matsuo2011} Matsuo, M., Ieda, J. I., Saitoh, E., Maekawa, S.: Phys. Rev. B {\bf 84}, 104410 (2011)

\bibitem{Sagnac} Sagnac, M. G.: C. R. Acad. Sci. (Paris) {\bf 157}, (1913)

\bibitem{Post} Post, E. J.: Rev. Mod. Phys. {\bf 39}, 475 (1967)

\bibitem{Barnett} Barnett, S. J.: Phys. Rev. {\bf 6}, 239 (1915)

\bibitem{Ono} Ono, M., Chudo, H., Harii, K., Okayasu, S., Matsuo, M., Ieda, J. I., Saitoh, E.: Phys. Rev. B {\bf 92}, 174424 (2015)

\bibitem{Einstein} Einstein, A., de Haas, W. J.: Verh. Dtsch. Phys. Ges. {\bf 17}, 152 (1915)

\bibitem{Mashhoon} Mashhoon, B.: Phys. Rev. Lett. {\bf 61}, 2639 (1988)

\bibitem{Fischer} Fischer, U. R., Schopohl, N.: Europhys. Lett. {\bf 54}, 502 (2001)

\bibitem{Viefers} Viefers, S.: J. Phys: Condens. Matter {\bf 20}, 123202 (2008)

\bibitem{Matsuo} Matsuo, M., Ieda, J. I., Saitoh, E., Maekawa, S.: Appl. Phys. Lett. {\bf 98}, 242501 (2011)

\bibitem{Schweikhard} Schweikhard, V., Coddington, I., Engels, P., Mogendorff, V. P., Cornell, E. A.: Phys. Rev. Lett. {\bf 92}, 040404 (2004)

\bibitem{Cooper} Cooper, N. R., Wilkin, N. K., Gunn, J. M. F.: Phys. Rev. Lett. {\bf 87}, 120405 (2001)

\bibitem{Bretin} Bretin, V., Stock, S., Seurin, Y., Dalibard, J.: Phys. Rev. Lett. {\bf 92}, 050403 (2004)

\bibitem{Shen} Shen, J. Q., He, S. L.: Phys. Rev. B {\bf 68}, 195421 (2003)

\bibitem{Lima} Lima, J. R., Brandão, J., Cunha, M. M., Moraes, F.: Eur. Phys. J. D {\bf 68}, 94 (2014)

\bibitem{Cooper2008} Cooper, N. R.: Adv. Phys. {\bf 57}, 539 (2008)

\bibitem{Lu} Lu, L. H., Li, Y. Q.: Phys. Rev. A {\bf 76}, 023410 (2007)

\bibitem{Werner} Werner, S. A., Staudenmann, J. L., Colella, R.: Phys. Rev. Lett. {\bf 42}, 1103 (1979)

\bibitem{Wang2018} Wang, B. Q., Long, Z. W., Long, C. Y., Wu, S. R.: Mod. Phys. Lett. A {\bf 33}, 1850025 (2018)

\bibitem{Santos} Santos, L. C. N., Barros, C. C.: Eur. Phys. J. C {\bf 78}, 13 (2018)

\bibitem{Hosseinpour2015} Hosseinpour, M., Hassanabadi, H.: Eur. Phys. J. Plus {\bf 130}, 236 (2015)

\bibitem{Castro} Castro, L. B.: Eur. Phys. J. C {\bf 76}, 61 (2016)

\bibitem{Dayi2018} Dayi, \"O. F., Yunt, E.: Ann. Phys. {\bf 390}, 143 (2018)

\bibitem{MatsuoPRL} Matsuo, M., Ieda, J. I., Saitoh, E., Maekawa, S.: Phys. Rev. Lett. {\bf 106}, 076601 (2011)

\bibitem{Anandan} Anandan, J.: Phys. Rev. D {\bf 24}, 338 (1981)

\bibitem{Zubkov} Zubkov, M. A.: Europhys. Lett. {\bf 121}, 47001 (2018)

\bibitem{Chernodub} Chernodub, M. N., Gongyo, S.: J. High Energy Phys. {\bf 2017}, 136 (2017)

\bibitem{Liu} Liu, Y., Zahed, I.: Phys. Rev. D {\bf 98}, 014017 (2018)

\bibitem{Chernodub2017} Chernodub, M. N., Gongyo, S.: Phys. Rev. D {\bf 96}, 096014 (2017)

\bibitem{Cavalcante} Cavalcante, E., Carvalho, J., Furtado, C.: Eur. Phys. J. Plus {\bf 131}, 288 (2016)

\bibitem{Gonzalez} Gonzalez, J., Guinea, F., Vozmediano, M. A. H.: Nucl. Phys. B {\bf 406}, 771 (1993)

\bibitem{Kolesnikov} Kolesnikov, D. V., Osipov, V. A.: Eur. Phys. J. B {\bf 49}, 465 (2006)
\bibitem{Cunha} Cunha, M. M., Brandão, J., Lima, J. R., Moraes, F.: Eur. Phys. J. B {\bf 88}, 288 (2015)

\bibitem{Gomes} Gomes, F. A., Bezerra, V. B., de Lima, J. R. F., Moraes, F. J. S.: Eur. Phys. J. B {\bf 92}, 41 (2019)

\bibitem{Ferkous} Ferkous, N., Bounames, A.: Phys. Lett. A {\bf 325}, 21 (2004)

\bibitem{Carvalho} Carvalho, J., Furtado, C., Moraes, F.: Phys. Rev. A {\bf 84}, 032109 (2011)

\bibitem{Andrade2014} Andrade, F. M., Silva, E. O.: Eur. Phys. J. C {\bf 74}, 3187 (2014)

\bibitem{Strange} Strange, P., Ryder, L. H.: Phys. Lett. A {\bf 380}, 3465 (2016)

\bibitem{B} Bakke, K.: Eur. Phys. J. Plus {\bf 127}, 82 (2012)

\bibitem{Ba} Bakke, K.: Gen. Rel. Gravit. {\bf 45}, 1847 (2013)

\bibitem{Vilenkin} A. Vilenkin, E. P. S. Shellard, {\it Cosmic strings and other topological defects} (Cambridge University Press, 2000)

\bibitem{Katanaev} Katanaev, M. O., Volovich, I. V.: Ann. Phys. {\bf 216}, 1 (1992)

\bibitem{Kleinert} Kleinert, H.:{\it Gauge Fields in Condensed Matter}, vol. 2 (World Scientific, Singapore, 1989)

\bibitem{Bakke2010} Bakke, K., Furtado, C.: Phys. Rev. D {\bf 82}, 084025 (2010)

\bibitem{Greiner} Greiner, W.: {\it Relativistic Quantum Mechanics: Wave Equations}, 3nd edn. (Springer, Berlin, 2000)

\bibitem{Aharonov} Aharonov, Y., Bohm, D.: Phys. Rev. {\bf 115}, 485 (1959)

\bibitem{Hehl} Hehl, F. W., Ni, W.-T.: Phys. Rev. D {\bf 42}, 2045 (1990)

\bibitem{Andrade} Andrade, F. M., Silva, E. O.: Europhys. Lett. {\bf 108}, 30003 (2014)

\bibitem{Villalba} Villalba, V. M., Maggiolo, A. R.: Eur. Phys. J. B {\bf 22}, 31 (2001)

\bibitem{Rubens} Oliveira, R. R. S., Sousa, M. F.: Braz. J. Phys. {\bf 49}, 315 (2019)

\bibitem{Abramowitz} Abramowitz, M., Stegum, I. A.:{\it Handbook of Mathematical Functions} (Dover Publications Inc., New York, 1965)

\bibitem{Vitoria} Vit\'oria, R. L. L., Bakke, K.: Int. J. Mod. Phys. D {\bf 27}, 1850005 (2018)

\bibitem{Mandal} Mandal, B. P., Verma, S.: Phys. Lett. A {\bf 374}, 1021 (2010)

\bibitem{Quimbay} Quimbay, C., Strange, P.: Quantum phase transition in the chirality of the (2+1)-dimensional Dirac oscillator, arXiv:1312.5251, (2013) [quant-ph]

\bibitem{Lamata} Lamata, L., Casanova, J., Gerritsma, R., Roos, C. F., Garc\'ia-Ripoll, J. J., Solano, E.: New J. Phys. {\bf 13}, 095003 (2011)

\bibitem{Schakel} Schakel, A. M.: Phys. Rev. D {\bf 43}, 1428 (1991)

\bibitem{Miransky} Miransky, V. A., Shovkovy, I. A.: Phys. Rep. {\bf 576}, 1 (2015)


\end{thebibliography}
\end{document}